\documentclass{article}

\usepackage{microtype}
\usepackage{multirow}
\usepackage{graphicx}
\usepackage{amsthm}
\usepackage{url}
\usepackage{fullpage}

\title{Packing identical simple polygons is NP-hard}
\author{Sarah R.~Allen \and John Iacono}
\date{Polytechnic Institute of New York University}

\newcommand{\prob}[1]{{\sc #1}}

\newtheorem{problem}{Problem}
\newtheorem{fact}{Fact}
\newtheorem{theorem}{Theorem}

\begin{document}

\maketitle

\begin{abstract}
Given a small polygon $S$, a big simple polygon $B$ and a positive integer $k$, it is shown to be NP-hard to determine whether $k$ copies of the small polygon (allowing translation and rotation) can be placed in the big polygon without overlap. Previous NP-hardness results were only known in the case where the big polygon is allowed to be non-simple. A novel reduction from \prob{Planar-Circuit-SAT} is presented where a small polygon is constructed to encode the entire circuit.
\end{abstract}

\section{Introduction}

Packing is a fundamental problem in computational geometry. In this paper we study the problem of packing multiple copies of a small object inside a big object:

\begin{problem}[Simple Polygon Packing] \label{p}
Given a \emph{small} simple polygon $S$, a \emph{big} simple polygon $B$, and a positive integer $k$, is it possible to place $k$ copies (allowing translation and rotation) of the small polygon inside the big polygon without overlap?
\end{problem}

This problem was heretofore neither known to be in P, nor in NP, nor to be NP-hard. Here we show Problem~\ref{p} is NP-hard. Several results are known about closely related variants of Problem~\ref{p}:

\begin{description}

\item[Multiple small polygons.] If there is not one small polygon, but rather a set of multiple small polygons, the problem is trivially NP-hard, even in one dimension.

\item[Big polygon is non-simple.] If the big polygon is non-simple (i.e.~it has holes), packing has been shown to be NP-hard even if the small polygon is a square \cite{DBLP:journals/ipl/FowlerPT81}.

\item[Small polygon is a square.] If the small polygon is a $2\times2$ square and all boundaries both the big and  the small polygons in a packing are restricted to be on the unit grid, then the problem was shown to be in NP \cite{DBLP:conf/cccg/El-KhechenDIO09}.

\item[Orthogonal rotation.] If orthogonal rotation is allowed, the problem is not known to be in NP, even if both the small and big polygons are rectangles. This is known as the \emph{pallet loading problem}, and appears as problem 55 on the \emph{Open Problems Project} \cite{opp}.

\end{description}

Our result is the first to establish the hardness of packing of multiple copies of a simple polygon inside another simple polygon. Previous reductions for related problems fall into two categories. In the case of multiple small polygons, a reduction from \prob{Knapsack} or \prob{Partition} is easy. In the case of having a nonsimple big polygon, the reduction in \cite{DBLP:journals/ipl/FowlerPT81} is from \prob{Planar-Circuit-SAT}. Such a reduction creates a big polygon which is essentially a drawing of the circuit, where the interior of the big polygon represents the wires and the gates, but where there are holes between all of the wires. Without the ability to literally create a big polygon that uses holes to create a circuit drawing, nothing was known. Our construction is also a reduction from \prob{Planar-Circuit-SAT}, but in a completely different manner. Previously the circuit was encoded in the big polygon; our big polygon is independent of all aspects of the circuit, other than the circuit size, while the circuit is encoded entirely in the small polygon. 

However, because our construction creates a small polygon which is nonconvex and nonconstant in size (the size is poynomial), there remains a range of open problems relating to the packing of identical polygons. 
The simplest such variation (most likely to be in P) would be: given as the big polygon an orthogonally convex simple polygon drawn on a unit grid, how many grid-aligned $2 \times 2$ unit squares can be packed? (This is a slightly easier variant of problem 56 on the \emph{Open Problems Project} \cite{opp}). The problem is known to be in P only if the big polygon is further restricted to be pyramidal \cite{d}. Harder variants of what are unknown to be in P or NP-hard include limiting the inside or outside polygons to be convex or of constant complexity.

\section{Planar circuit satisfiability on a grid}

Our reduction is from a problem we call \prob{Planar-Grid-SAT}:\, which is designed for the ease of presentation of the reduction of \S\ref{reduction}:

\begin{problem}[Planar-Grid-SAT]
The input is an integer $n$ and a classification of the edges on an $n\times m$ grid such 
that some of the edges of the grid are designated to be \emph{wires}, other edges are designated to be \emph{inverters}, and some pairs of adjacent horizontal edges where the rightmost vertex of the three has an odd $x$ coordinate are called \emph{\textsc{And} gates}.
Those edges which are unclassified are referred to as \emph{don't cares}. The input is \emph{satisfiable} if and only if there is an assignment of \textsc{True} or \textsc{False} to every vertex on the grid such that 
\begin{itemize}
\item vertices connected by a wire have the same truth value,
\item vertices connected by an inverter have different truth values,
\item For three vertices connected by an \textsc{And} gate, the truth value of the right vertex is the logical \textsc{And} of the middle and left vertex, and
\item The upper-left vertex has a truth value of \textsc{True}.
\end{itemize}

 \end{problem}

The NP-hardness of determining whether an instance of \prob{Planar-Grid-SAT} is satisfiable follows from the NP-hardness of \prob{Planar-Circuit-SAT} and the fact that any planar graph can be drawn on a polynomial size grid \cite{DBLP:journals/combinatorica/FraysseixPP90}; the location restriction of the {\sc And} gates is trivial to enforce through expansion of the grid size by a constant factor. \prob{Planar-Circuit-SAT} itself is known to be NP-hard via a well-known reduction from \prob{Circuit-SAT}, where overlapping wires in a non-planar drawing of a circuit can be implemented in an equivalent planar circuit through the appropriate use of three \textsc{Xor} gates (for an illustration of this construction see, for example, \S3.2.3 of \cite{psat}).

\section{Reduction}\label{reduction}

Due to the complexity of the reduction, it is presented in several phases of increasing complexity. In the first three phases (\S\ref{p1}-\ref{p3}) we simply present big and small polygons, such that for a particular $k$ the packing of $k$ small polygons inside the big one is unique. In \S\ref{p4} we add a small degree of discrete flexibility to how the small polygons can be packed; each small polygon will represent a vertex in a instance of \prob{Planar-Grid-Sat}, and this flexibility represents the truth value of that vertex. Finally, in \S\ref{p4} we restrict the flexibility of the positions of the small polygons granted in the previous phase to that which represents truth assignments to the vertices which are allowable in the given instance of \prob{Planar-Grid-Sat}.

\subsection{Squares in a rectangle}\label{p1}

We begin with the obvious:

\begin{fact}

Given a $n\times m$ rectangle as the big polygon, and a unit square as the small polygon, there is a unique way to pack $nm$ copies of the small polygon in the big polygon.
\end{fact}

\subsection{Protrusions, inclusions, and a progression of whitespace}\label{p2}

\begin{figure}
\begin{center}
{\includegraphics[width=5.5in, height=6.5in, trim=40 0 120 200, clip=true]{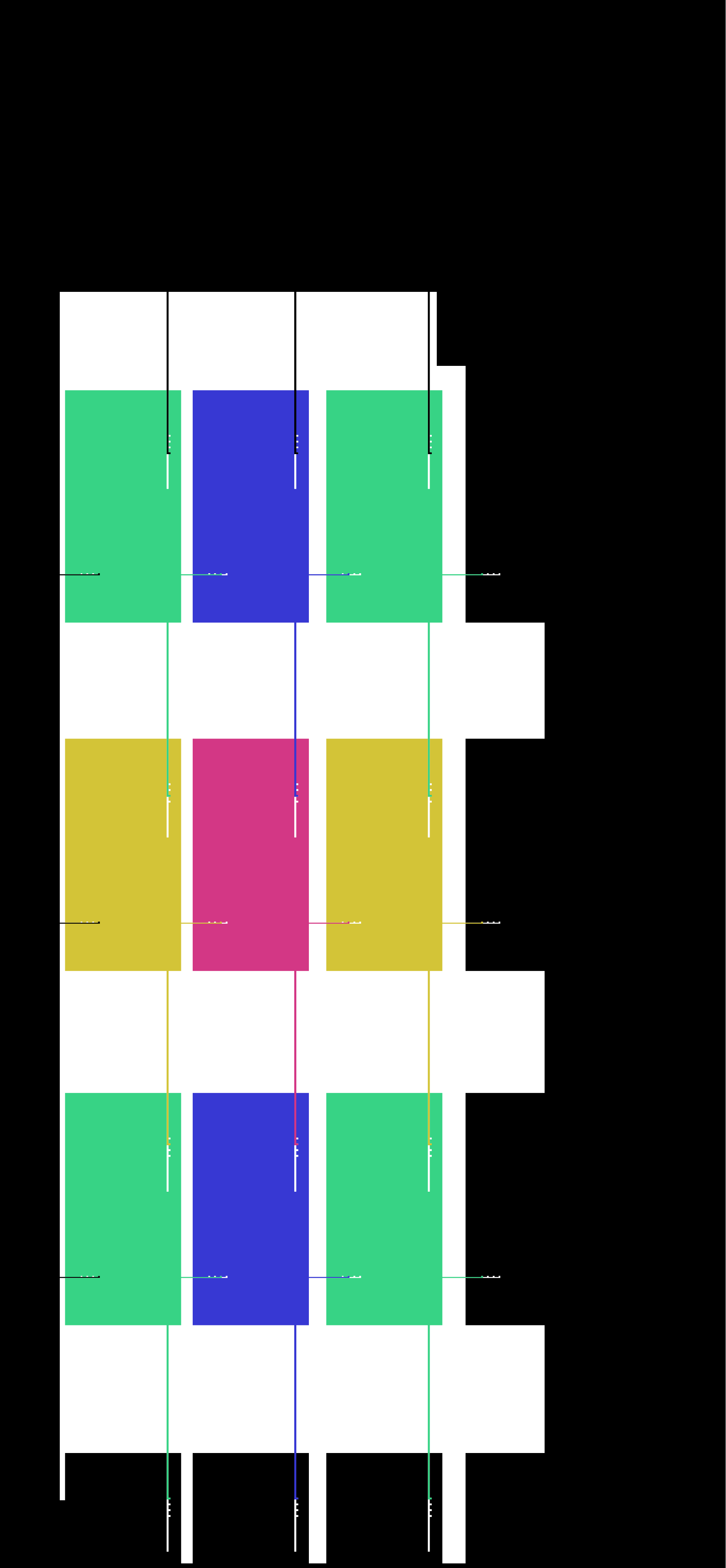}}
\caption{Illustration of the graduated grid packing. The colored polygons are all identical, and the interior boundary of the black region is the big polygon. For a closeup of the horizontal protrusion, see Figure~\ref{f:hibig}.  }
\end{center}
\label{fig:grad}
\end{figure}

In this phase we maintain the unique packing, but add some whitespace to the solution; see Figure~\ref{fig:grad}. Additionally, the small polygon, while still roughly square-shaped has two \emph{protrusions}, horizontal and vertical, and two matching \emph{inclusions}. The big polygon also now has inclusions and protrusions which force the location of where each row and column of the squares line up. The whitespace is chosen so that the distance between each row and column increases linearly as one proceeds down and to the right. Additionally, the total whitespace is chosen to have a total area of less than 1, thus making it obvious that the packing of $nm$ small polygons is unique.

Observe that each inclusion has $n$ \emph{notches}. Also observe that by construction, in a given small polygon, the vertical notch which the protrusion from above occupies is a graphical representation of the $y$ coordinate of the small polygon in the grid of packed small polygons. Symmetrically, the horizontal notch occupied is a graphical representation of the horizontal position within the grid; see see Figure~\ref{fig:grad}.

This is a good first step of making a polygon ``aware'' of its location, however, we would like to be able to make the notch a neighboring protrusion occupies a function of the small polygon's horizontal \textbf{and} vertical positions; currently the notch is a function of the horizontal \textbf{or} the vertical position.

\subsection{The nailer}\label{p3}

\begin{figure}
\begin{center}
{\includegraphics[width=5.5in, height=6.5in, trim=40 0 120 200, clip=true]{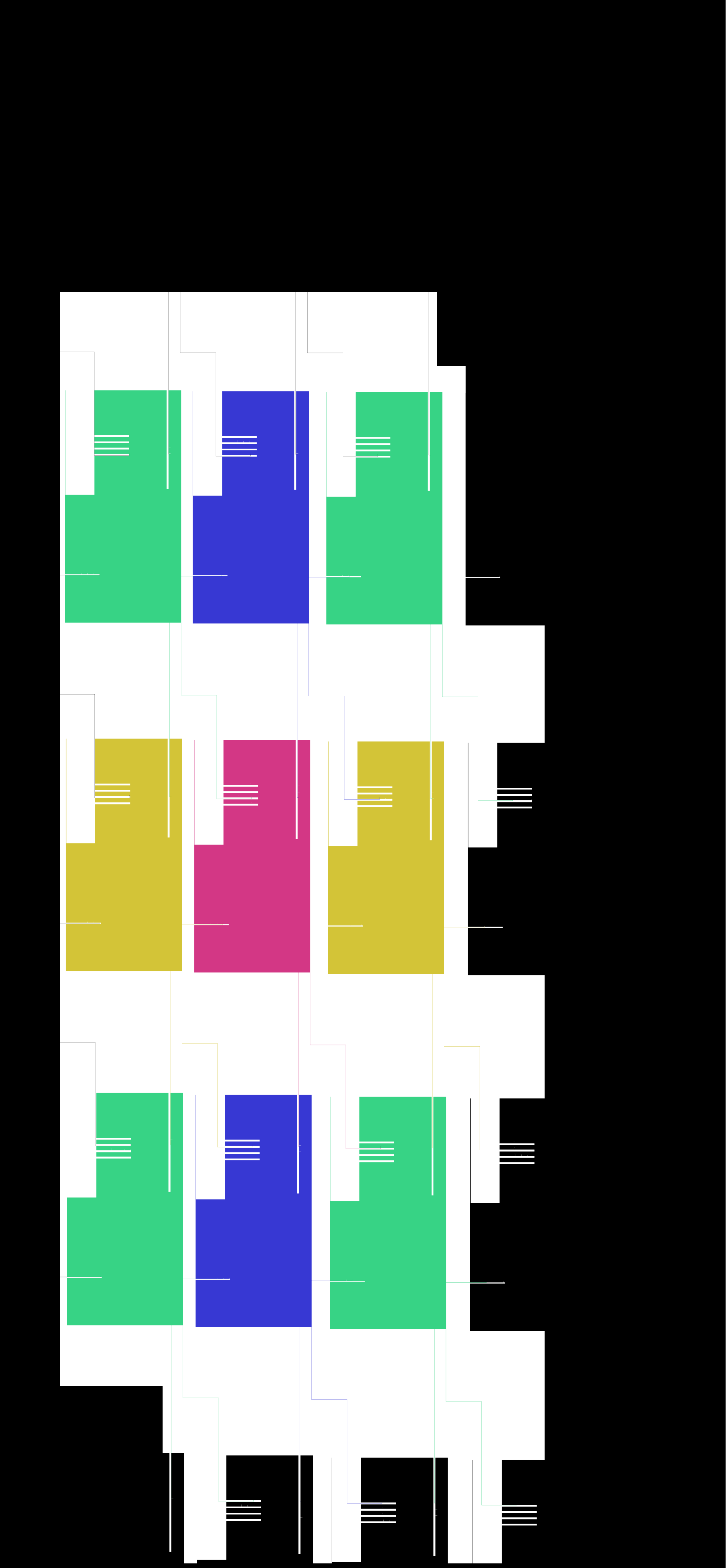}}
\end{center}
\caption{Illustration of the nailer and the nail. Notice how in each small polygon the nail nails the nailer in a different notch. See Figure~\ref{f:hi} for a closeup of the nailer. (Note that in a PDF viewer, the reader is encouraged to zoom in to explore the details)}
\label{fig:p4}
\end{figure}

In this phase the location of each small polygon in a packing of $nm$ of them remains exactly the same as in the previous one. The only change is that a protrusion (called the \emph{nail}) is added heading down and to the right from the bottom right corner of each small polygon, and a matching inclusion (called the \emph{nailer}) is built on the upper left portion of each polygon so that the nail never overlaps a neighboring small polygon. Observe that the nailer has $nm$ notches, arranged in an $m \times n$ grid, and for each small polygon's nailer, the notch occupied by a neighboring polygon's nail is unique for each small polygon and is a graphical representation of both the horizontal and vertical position of the small polygon in the grid of packed small polygons. Thus, we have for the first time have a local part of each small polygon's packing which is uniquely determined by its position, both horizontal and vertical. This will be exploited further in the next step.

\subsection{Refining the nailer---microshifts and subnotches}\label{p35}

Recall that since \S\ref{p2}, the horizontal notch occupied is the horizontal position and the vertical notch occupied is the vertical position. In this phase we make several changes:

%OUT OF ORDER WITH THE CONSTUCTION
%\begin{figure}
%{\includegraphics[width=5.5in, height=5in, trim=40 0 120 200, clip=true]{polygon/phase3.pdf}}
%\caption{Here the protrusions have been shrunk, and the notches subdivided into micronotches. See Figure~\ref{f:hibig} for a closeup of the micronotches. At this point the polygons could be packed using any micronotch in a group; this flexibility is removed in the next step through the introduction of the nailer. }
%\label{fig:p3}
%\end{figure}

\begin{figure}
\begin{center}
{\includegraphics[width=5in, trim=150 520 1730 535, clip=true]{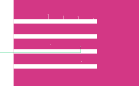}}
\caption{Nailer closeup. The notch that the nail will end up in is uniquely determined by the position of the small polygon. Observe how the heights and horizontal alignment of each notch is different, this ensures the desired micronotch placement.}
\end{center}
\label{f:nail}
\end{figure}

\begin{itemize}

\item We refine each of the $n$ horizontal notches in the horizontal inclusion into $m$ \emph{micronotches}, where $m$ represents the number of vertical notches. Thus the horizontal inclusion now has $nm$ micronotches, which are arranged in $n$ \emph{groups} of $m$. See Figure~\ref{fig:p4} for an illustration of this and Figure~\ref{f:hi} for a closeup of the nailer. Formerly, the occupied notch represented the horizontal location; now the micronotch group uniquely determines the horizontal coordinate, and the micronotch within the group will uniquely determine the vertical coordinate within the group. Symmetric changes are performed in the vertical protrusion and inclusions.

\item We shrink the horizontal and vertical protrusions so that they fit in the micronotches.

\item We make small shifts in the inclusions of the nailer, so that the horizontal and vertical probes uniquely fit in the appropriate micronotch. For an illustration of the resultant nailer see Figure~\ref{f:nail}.

\end{itemize}

\begin{figure}
\begin{center}
{\includegraphics[width=5in, trim=120 442 340 632, clip=true]{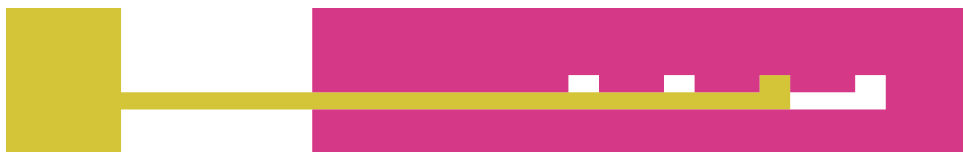}}
\caption{Horizontal inclusion closeup.}
\end{center}
\label{f:hibig}
\end{figure}

\begin{figure}
\begin{center}
{\includegraphics[width=5in, trim=120 442 1760 635, clip=true]{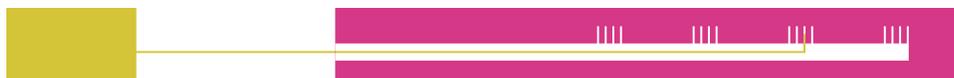}}
\caption{Horizontal inclusion closeup, showing the micronotches arranged in groups. Compare to Figure~\ref{f:hibig}}
\end{center}
\label{f:hi}
\end{figure}

The resulting packing of $nm$ small polygons remains unique. The reason is that while the horizontal and vertical protrusions have more wiggle room than in the previous phases, this wiggle room is chosen to be less than the distance between the notches in the nailer, thus the nail is nailed in the nailer in a unique way which fixes the position uniquely and establishes the correct positioning of the protrusions in the micronotches.

\subsection{Adding states}\label{p4}

\begin{figure}
\begin{center}
{\includegraphics[width=5.5in, height=6.5in, trim=40 0 120 200, clip=true]{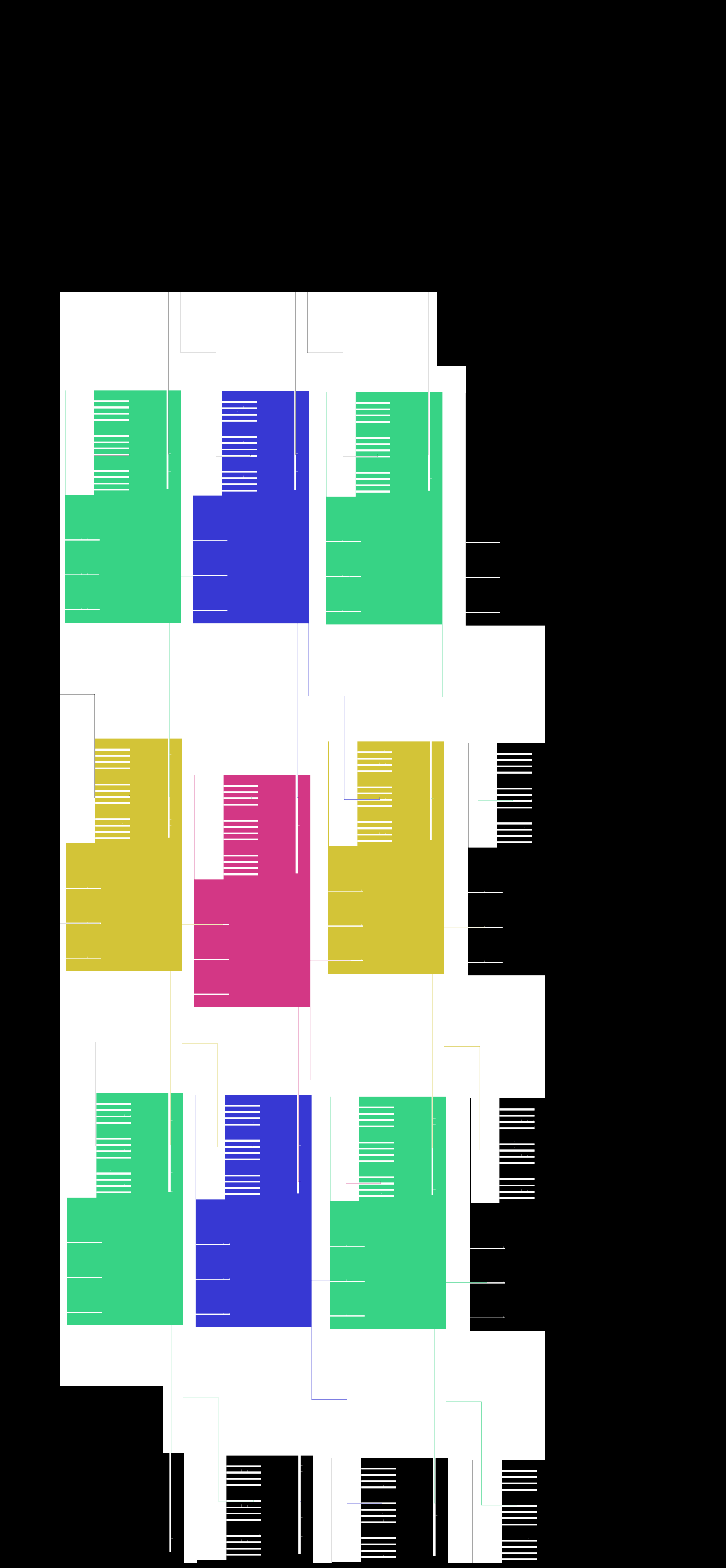}}

\caption{To introduce the concept of states, the horizontal and vertical inclusions as well as the nailer are copied several times. In this figure we illustrate three copies. We allow the small polygons to be packed either next to the polygon to their left, or above or below; this will determine which of the shifted inclusions is used. This is a simplification of the construction of the text, where the small polygons could be shifted in six ways, with a ratio of 31 between the large and the small shift, and the number copies of the inclusions being 63 instead of three. Also note that for ease of viewing the whitespace and other visual elements have been exaggerated dramatically; the total whitespace should be smaller than a single small polygon, and all inclusions should be much smaller and horizontally distant from the exterior of the small polygons.}
\end{center}
\label{fp5}
\end{figure}

The last phase left us with a construction whereby the packing of $nm$ small polygons was unique, and that for each of the small polygons the micronotch occupied was unique. In order to implement a circuit, establishing this  this location awareness among identical objects is crucial, but now we must add a degree of flexibility which will allow a discrete degree of flexibility on the location where each small polygon is packed. The \emph{state} of a small polygon in a packing will be represented by a vertical shift. The size of this shift is relatively small compared to the size of a unit square, and relatively large compared to the nailer. We modify the construction of the previous phase to create several vertically translated copies of the horizontal and vertical inclusions and nailer. This is illustrated in Figure~\ref{fp5}.

To have a small polygon represent a node in the the circuit in a given instance of \prob{Planar-Grid-SAT}, we need to know if a given small polygon represents a true or a false state. For technical reasons explained below we also need to know if the polygon is currently in an even or odd numbered column. Finally, since the right node of an {\sc And} gate in an instance of \prob{Planar-Grid-SAT} is a function of the truth states of not just the neighbor to the left, but two neighbors to the left we will need to encode in the state the truth value of the left neighbor; for simplicity we only encode this for even columns. This gives the six states listed in in Table~\ref{table:states}. We assign each state a shift value according to an exponential progression, with the largest shift being 31 times the smallest. The reason this exponential progression is chosen is so that if two adjacent small polygons are in different states, the horizontal difference between them will uniquely determine their states. Observe that if there is no horizontal difference between adjacent small polygons, we know that they are in the same state, but we do not know which state that is; this is the reason for also encoding the columnar parity, as this means that having two horizontally adjacent small polygons in the same state is forbidden.

Thus, in a given small polygon, the horizontal protrusion from the left neighbor will go into a single micronotch (vertical coordinate) in a group of notches (horizontal coordinate) in a copy of the horizontal protrusion which indicates the shift difference between the given small polygon and the polygon to the left. As this shift difference encodes the states of both the given polygon and the polygon to the left, the exact micronotch occupied by the protrusion of the polygon to the left will be a unique function of the location of the given small polygon, its truth value, the truth value of the polygon to the left, and the truth value of the polygon two to the left (if the given polygon is in a odd column). Based on all of this information, it can be determined if the micronotch occupied represents a truth assignment that is consistent with the given instance of \prob{Planar-Grid-SAT}. For example, if there were a wire connecting the node representing the current polygon to the node representing the one to the left, then we want to allow the grid packing of $nm$ small polygons only if the two polygons have the same truth value. This can easily be done by removing the micronotches that do not correspond to this condition. Our construction ensures that by keeping or removing micronotches in the small polygon, we can allow a packing of $nm$ small polygons if and only if the circuit is satisfied. Table~\ref{table:h} indicates how we can choose which horizontal micronotches to keep as a function of the circuit element being represented. Programming the vertical micronotches is slightly easier, since all {\sc And} gates are represented horizontally; Table~\ref{table:v} has the relevant information for vertical micronotch selection.

The large polygon is, in essence, created in the same manner as an enclosing ring of small polygons; the micronotches are encoded such that the interface between the big polyon and small polygons pack inside is a don't care, except for at the upper-true where a {\sc True} condition is enforced. This corresponds to enforcing that the circuit be satisfied.

Thus we can summarize the result of the construction in this theorem:

\begin{theorem}
Given an instance of \prob{Planar-Grid-SAT} on an $n\times m$ circuit, 
the construction of this section yields a small polygon $S$ and a large polygon $B$ such that
$nm$ small polygons can be packed in the large polygon in a grid like-manner if and only if their shifts represent truth values which are consistent with the circuit diagram.
\end{theorem}

\begin{table}
\begin{center}
\begin{tabular}{|ccc|} \hline
Shift Amount & Abbreviation & Description of State\\ \hline
0 & OT & Odd column, truth value of {\sc True}\\
1 & OF & Odd column, truth value of {\sc False}\\
3 & ETT & Even column, truth value of {\sc True}, polygon to the left is {\sc True}\\
7 & ETF & Even column, truth value of {\sc False}, polygon to the left is {\sc True}\\
15 & EFT & Even column, truth value of {\sc True}, polygon to the left is {\sc False}\\
31 & EFF & Even column, truth value of {\sc False}, polygon to the left is {\sc False} \\ \hline
\end{tabular}
\end{center}
\caption{Description of states}
\label{table:states}
\end{table}

\begin{table}
\begin{center}
\begin{tabular}{|c|c|c|c|c|c|c|c|c|c|} \hline
\multicolumn{9}{|c|}{\large Horizontal Micronotch Programming} \\ \hline
\multirow{2}{*}{Difference}& \multirow{2}{*}{Transition}  & \multicolumn{3}{c|}{Even column} & \multicolumn{4}{c|}{Odd column} \\ 
& & Wire & Inverter & Don't care & Wire & Inverter & AND &Don't care\\ \hline
-31& EFF $\rightarrow$ OT & & & & &$\circ$ & &$\circ$ \\
-30& EFF $\rightarrow$ OF & &&  &$\circ$ & &  $\circ$&$\circ$ \\
-15 & EFT $\rightarrow$ OT & & &&$\circ$ & &&$\circ$  \\
-14 & EFT $\rightarrow$ OF & && & &$\circ$ & $\circ$ &$\circ$\\
-7 & ETF $\rightarrow$ OT & & && &$\circ$ &&$\circ$  \\
-6 & ETF $\rightarrow$ OF & && &$\circ$ & & $\circ$&$\circ$ \\
-3 & ETT $\rightarrow$ OT & && &$\circ$ & &$\circ$ &$\circ$ \\
-2 & ETT $\rightarrow$ OF & && & &$\circ$ &  &$\circ$\\
2& OF $\rightarrow$ ETT  & &$\circ$ &$\circ$ & & &  &\\
3& OT $\rightarrow$ ETT  &$\circ$ & &$\circ$ & & &  &\\
6 & OF $\rightarrow$ ETF &$\circ$ & &$\circ$ & & &&  \\
7 & OT $\rightarrow$ ETF & &$\circ$ &$\circ$ & & & & \\
14 & OF $\rightarrow$ EFT & &$\circ$ &$\circ$ & & &&  \\
15 & OT $\rightarrow$ EFT &$\circ$ & & $\circ$ && & & \\
30 & OF $\rightarrow$ EFF &$\circ$ & &$\circ$ & & & & \\
31 & OT $\rightarrow$ EFF & &$\circ$ &$\circ$ & & & & \\ \hline
\end{tabular}
\end{center}
\caption{Horizontal micronotch programming. This programs the relationship between horizontally adjacent blocks as a function of the parity of their column and the part of the circuit diagram they are representing. Only transitions from even to odd or vice-versa are allowed as the column parity of adjacant columns must be different. Micronotches should only be present when they represent a node in the circuit with the given difference which in the given instance of \prob{Planar-Grid-SAT} has a logical element allowed by the programming function.  Allowed micronotches are designated by a $\circ$ in the above table. }
\label{table:h}
\end{table}

\begin{table}
\begin{center}
\begin{tabular}{|c|c|c|c|c|c|c|c|c|} \hline
\multicolumn{8}{|c|}{\large Vertical Micronotch Programming} \\ \hline
\multirow{2}{*}{Difference}& \multirow{3}{*}{Transition}  & \multicolumn{3}{c|}{Even column} & \multicolumn{3}{c|}{Odd column} \\ 
& & Wire & Inverter & Don't care & Wire & Inverter & Don't care\\ \hline
-28 & EFF $\rightarrow$ ETT & & $\circ$& $\circ$& &  & \\
-24 & EFF $\rightarrow$ ETF &$\circ$ && $\circ$ & &   & \\
-16 & EFF $\rightarrow$ EFT & &$\circ$& $\circ$ & &   &\\
-12 & EFT $\rightarrow$ ETT &$\circ$ && $\circ$ & &  &\\
-8 & EFT $\rightarrow$ ETF & &$\circ$& $\circ$ & &   &\\
-4 & ETF $\rightarrow$ ETT & &$\circ$& $\circ$ & &  &\\
-1 & OF $\rightarrow$ OT & & & &  & $\circ$ & $\circ$ \\
0 & \fbox{\parbox{1in}{ \tiny
EFF$\rightarrow$EFF
EFT$\rightarrow$EFT
ETF$\rightarrow$ETF
ETT$\rightarrow$ETT
OF$\rightarrow$OF
OT$\rightarrow$OT
}} & $\circ$ & & $\circ$ & $\circ$& & $\circ$ \\
1& OT $\rightarrow$ OF  & &&  & &  $\circ$& $\circ$ \\
4& ETT $\rightarrow$ ETF  & & $\circ$& $\circ$& & &  \\
8 &ETF $\rightarrow$ EFT & & $\circ$& $\circ$& &   &\\
12 & ETT $\rightarrow$ EFT & $\circ$&& $\circ$ & & &  \\
16 & EFT $\rightarrow$ EFF & & $\circ$& $\circ$& &  & \\
24 & ETF $\rightarrow$ EFF &$\circ$ && $\circ$ & &  & \\
28 & ETT $\rightarrow$ EFF & &$\circ$& $\circ$ & &  & \\ \hline
\end{tabular}
\end{center}
\caption{Vertical micronotch programming. This programs the relationship between vertically adjacent blocks as a function of the parity of their column and the part of the circuit diagram they are representing. }
\label{table:v}
\end{table}

\begin{figure}
\begin{center}
{\includegraphics[width=5in, height=2in, trim=50 350 1600 500, clip=true]{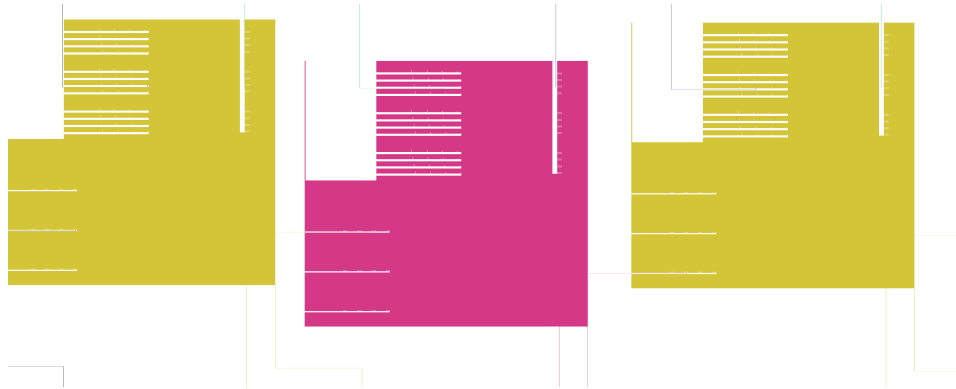}}
\caption{States. This closeup shows the middle polygon is in a different state from its neighbors.}
\end{center}
\end{figure}

\section{Acknowledgments}
The work here was partially completed at the 15th Korean Workshop on Computational Geometry held at Carleton University in Ottawa, Ontario, Canada in August of 2012. We would like to thank the organizers of this event, Prosenjit Bose, Vida Dujmovi\'{c}, Anil Maheshwari, Pat Morin, and Michiel Smid. The research of the first author was supported by NSF grant CCF-0830734, and research of the second author is supported by NSF grant CCF-1018370.

\bibliography{refs}

\end{document}